\documentclass{aa}  

\usepackage{graphicx}
\usepackage{xcolor}
\usepackage{txfonts}
\usepackage{hyperref}
\hypersetup{colorlinks,allcolors=black}
\begin{document}

   \title{Did the Milky Way just light up?\\ The recent star formation history of the Galactic disc}
     \author{Eleonora Zari, 
          \inst{1}
          Neige Frankel,
          \inst{2}
          \and
          Hans-Walter Rix \inst{1}
          }

   \institute{Max-Planck-Institut f\"ur Astronomie, K\"onigstuhl 17 D-69117 Heidelberg, Germany\\
         \and
             Canadian Institute for Theoretical Astrophysics, University of Toronto, 60 St. George Street, Toronto, ON M5S 3H8, Canada\\
             }

 
\abstract{We map the stellar age distribution ($\lesssim 1$~Gyr) across a 6kpc$\,\times\,$6kpc area of the Galactic disc, in order to constrain our Galaxy's recent star-formation history. Our modelling draws on the sample of Zari et al. (2021) that encompasses all presumed disc OBA stars ($\sim 500,000$ sources) with $G<16$. To be less sensitive to reddening, we do not forward model the detailed CMD distribution of these stars, but instead the K-band absolute magnitude distribution, $n(M_K)$ among stars with $M_K<0$ and $T_{\mathrm{eff}} > 7000$~K at a certain positions $\vec{x}$ in the disc as a step function with five age bins, $b(\tau~|~\vec{x}, \vec{
\alpha})$, logarithmically-spaced in age from $\tau = 5$~Myr to $\tau \sim 1$~Gyr.   Given a set of isochrones and a Kroupa (2001) initial  mass function, we sample $b(\tau\,|\,\vec{x}, \vec{
\alpha})$ to maximise the likelihood of the data $n(M_K\,|\,\vec{x}, \vec{\alpha})$, accounting for the selection function. This results in a set of mono-age stellar density maps across a sizeable portion of the Galactic disc. These maps show that some, but not all, spiral arms are reflected in overdensities of stars younger than 500~Myr. 
The maps of the youngest stars ($<10$~Myr) trace major star forming regions. The maps at all ages exhibit an outward density gradient and  distinct spiral-like spatial structure, which is qualitatively similar on large scales among the five age bins.  When summing over the maps' area and extrapolating to the whole disc, we find an effective star-formation rate over the last 10~Myr of $\approx 3.3 \mathrm{M_{\odot}/yr}$, higher than previously published estimates that had not accounted for unresolved binaries. Remarkably, our stellar age distribution implies that the star-formation rate has been three times lower throughout most of the last Gyr, having risen distinctly only in the very recent past. Finally we use TNG50 simulations to explore how justified the common identification of local \emph{age distribution} with global \emph{ star formation history} is: we find that the global star-formation rate at a given radius in Milky-Way-like galaxies is approximated within a factor of $\sim 1.5$ by the young age distribution within a 6kpc$\,\times\,$6kpc area near $R_\odot$.}

\date{Received -; accepted -}
\keywords{stars: early-type; Galaxy: structure; Galaxy: disk;}
\titlerunning{SFH Galactic disc}
\authorrunning{Zari, Frankel, Rix}
\maketitle

\section{Introduction}
Galaxies are fundamentally shaped by their star formation history. At any given epoch, the star formation rate (SFR) determines the population of massive, short lived stars that regulate and change the composition and energetics of ISM. The SFR sets the rate at which the ISM is dispersed through galactic winds or locked away into long-lived, low-mass stars. It correlates with the nuclear activity of galaxies and the growth of stellar mass throughout cosmic time \citep{Madau1996, Bouwens2011}. Finally, the SFR is a fundamental input and observational constraint for galaxy evolution models \citep{Chiappini1997, Kauffmann2000, deRossi2009}, again because it sets both the level of massive star feedback and the rate at which gas and dust are depleted.

In external unresolved galaxies the SFR is most often determined via a wide range of proxy measurements
\citep[see e.g.][]{Kennicutt2012}. In the Milky Way, one can -- at least in principle -- count up the young stars. Low-mass stars can be discerned as being young if they have not yet reached the main sequence; massive stars can be discerned as young by their CMD position, or simply by the fact that they are short-lived. But not all stars can readily be recognised as young. Therefore, any such counting exercise must account for the fraction of the stellar mass function enclosed in such a census.

For nearby ($d \lesssim 500$~pc) molecular clouds in the Milky Way, nearly complete lists of young stellar objects (YSOs) are available, whose ages ($t$) can be derived by measuring their CMD positions or infrared excesses. The SFR  is therefore given by $ \langle \dot{M}_{*}\rangle = N_{YSO}~ \langle M_{YSO}\rangle /t$, where $<M_*>$ is the mean mass of YSOs and is usually assumed to be $\langle M_*\rangle = 0.5 \, \mathrm{M_{\odot}}$ \citep[see e.g.][]{Lada2010}. In the local group, analogous SFR estimates have been based on the CMD position of massive luminous stars.  For example, \cite{Williams2011}  have applied CMD fitting techniques to reconstruct spatially resolved maps of the stellar age distribution in nearby galaxies. 

The ``current'' SFR of the Milky Way has been estimated by various groups over the last decade, e.g. \cite{Robitaille2010}, \cite{Chomiuk2011}, and \cite{Licquia2015}.
Specifically, \cite{Robitaille2010} used the census of young stellar objects (YSOs) from \citet[][hereafter R08]{Robitaille2008} to construct a population synthesis model for YSOs in the Galaxy, applied the same observational constraints as for the R08 census,  and varied the model SFR such that the number of detected synthetic YSOs matched the observed number. They inferred a global SFR (across the entire Milky Way) of 0.68-1.45 $\mathrm{M_{\odot} \, yr^{-1}}$. 
\cite{Chomiuk2011} normalised various estimates for the Galactic SFR to the same initial mass function (IMF) and population synthesis models across methods, and estimated $<\dot{M_*}> = 1.9 \pm0.4~\mathrm{M_{\odot}}/yr$.
\cite{Licquia2015} revisited this work and estimated $<\dot{M_*}> = 1.65\pm0.19~\mathrm{M_{\odot}}/yr$ by combining previous measurements from the literature using a hierarchical Bayesian statistical method.

A conceptual limitation of such analyses is the identification of a measured \emph{age distribution} of stars in a Galactic neighbourhood with the local or global \emph{star formation history} of the Milky Way, for two reasons. First, young stars that are now seen in a modest volume (say, a few 100 pc) corotating with the Sun were not necessarily born within that volume. Second, we know from external disk galaxies that the instantaneous SRF varies dramatically as a function of position within a galaxy \citep[see e.g. Fig.~1 in][for a recent illustration]{Kreckel2018}; therefore, the extrapolation of a local census of young stars to a galaxy-wide SFR may be far off. To acknowledge this crucial distinction, but not completely discard traditional terminology, we use the term \emph{effective star formation rate} for the quantity that is naively inferred from the age distribution. In addition, the SFR of galaxies may vary on short time scales, so that any estimate of $\dot{\mathrm{M}}_{*}\sim \mathrm{M}_*(<\Delta t)/\Delta t$ will depend on the age range $\Delta t$ of the young stars considered.

In this study, we set out to model and determine the age distribution over a substantive fraction ($6\times6$~kpc around the Sun) of the Galactic disc, using upper main sequence, i.e. OB(A), stars. To do this we devise and apply a non-parametric model\footnote{a model for which we do not assume a specific functional form, but have parameters for} for the age distribution of stars, which predicts their absolute magnitude distribution via isochrones. These predictions are then compared to OB(A) stars across many spatial patches in the Galactic disc, resulting in age-resolved spatial maps of stars younger than 1~Gyr. On that basis we explore to which extent the modelled \emph{age distribution} and its \emph{effective SFR} reflect the actual disc star-formation history of the Milky Way.  
Recently, \cite{RuizLara2020} have made a comprehensive estimate of the age distribution of Galactic disc stars within 2~kpc  by modelling \textit{Gaia} DR2 colour-magnitude diagrams. However, their focus was on 0.5-10~Gyr timescales as opposed to the younger populations that we study here.
Our analysis draws in the sample of massive stars from \cite{Zari2021}, which we describe in Section \ref{sec:data}. We lay out our methodology in Section \ref{sec:model}. Our results, presented in Section \ref{sec:results}, show a rapid increase in star formation rate compared to previous years, which could be related to the presence of numerous massive star-forming regions in the area considered. We discuss our results and summarise them in Section \ref{sec:discussion}.

\begin{figure*}[h]
    \centering
    \includegraphics[width=\hsize]{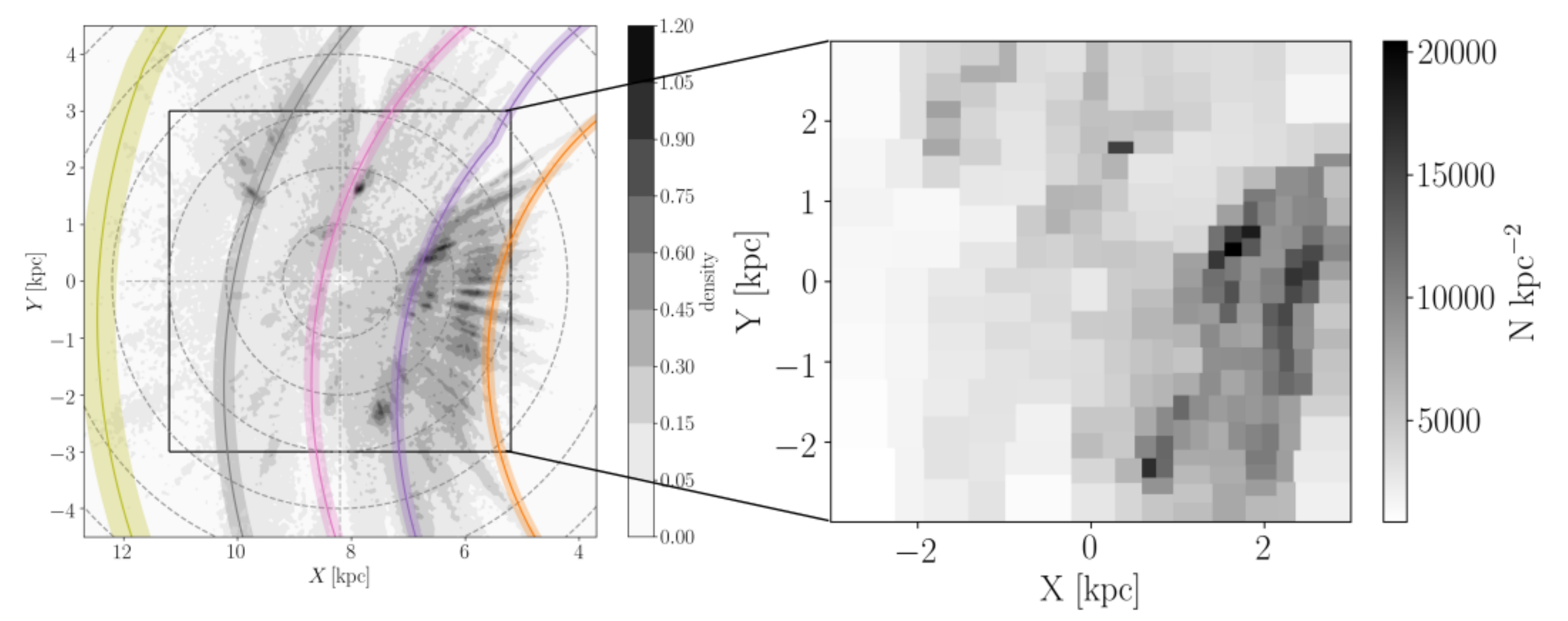}
    \caption{Density distribution of the filtered sample of massive hot stars from \cite{Zari2021}. In the left panel, the  coloured lines represent the spiral arm locations derived by \cite{Reid2019}. The right panel highlights the $6\times6$~kpc region considered in this study. Such region is divided in "spaxels" of different sizes, each containing the same number of sources. In both panels, the Sun is in  $X,Y = (0,0)$.
   }
    \label{fig:density}
\end{figure*}

\section{Data: Luminous Young Disc Stars}\label{sec:data}
The sample that we used in this study consists of $\sim 500,000$ OB(A) stars in the "filtered" sample presented in \cite[][hereafter Z21]{Zari2021}. The sample was selected by combining Gaia EDR3 
photometry and astrometry and 2MASS photometry by applying the following criteria. First, Z21 selected sources brighter than $G = 16$~mag and with absolute magnitude in the 2MASS $K_s$ band $M_K < 0$~mag, which roughly corresponds to a late B-type stars. Then, they applied several cuts in colour-colour space to exclude bright red giant- and asymptotic giant-branch stars. To clean the sample from spurious astrometric solutions, they removed sources with astrometric fidelity $< 0.9$ \citep{Rybizki2022}. Finally, they selected sources in the Galactic disc ($|z| < 300$ pc) and with small vertical velocities $v_z$.

Z21's sample extends to distances of $\sim$5~kpc from the Sun (Fig. \ref{fig:density}, left). However, in this study we focus on a 6kpc$\times$6kpc region of the Galactic disc, centred on the Sun, shown in Fig. \ref{fig:density} (right), where we can reasonably assume that our sample, as selected in Eq. \ref{eq:selfunc},  is complete.
To derive our spatially resolved SFR map of the Galactic disc, we divide our volume into 256 "spaxels" of different sizes, each containing the same number of stars ($\sim1000$). We model the magnitude distribution of the sources in each spaxel, as described in the next Section.

\section{Modelling the Stellar Age Distribution }\label{sec:model}
Ideally, we would like to model the observed colour-magnitude distribution of our sample stars in various spaxels to constrain their age distribution. However, this would require detailed star-by-star star dereddening, which would be difficult if not impossible \citep[but see e.g.][]{RuizLara2020}. Therefore, we resort to modelling a more restricted set of observables, the distribution of absolute magnitudes $M_K$ for stars that are luminous ($M_K<0$) and (intrinsically) bluer than the red-giant branch; this latter colour criterion can be satisfied even in the presence of extensive reddening. These stars should all be massive young main sequence or evolved stars (Z21).

Specifically, we chose, as a statistic $n\left( M_K~|~\vec{x}, \vec{\alpha} \right)$, the absolute K-band magnitude distribution of sample members, which we take to be a function of position in the Galaxy $\vec{x}$. And we assume that at each $\vec{x}$ there is an underlying age or birthrate distribution, $b\bigl ( \tau~|~\vec{x}, \vec{\alpha} \bigr)$, whose temporal dependence is described by a set of parameters $\vec{\alpha}$, specified in Eq.\ref{eq:stepfunction}. The expected observed distribution in absolute magnitudes then depends on stellar evolution, the initial mass function, the birthrate distribution and the observational selection function:
\begin{align} 
\label{eq:basic-prediction}
    n\left( M_K~|~\vec{x}, \vec{\alpha} \right)  = 
    \int d\tau~\int dm_0~\int dT_{\mathrm{eff}} \ \ \times \nonumber \\ 
    \ \ \ \ \ S_c(M_K, T_{\mathrm{eff}})\ \ 
    p(M_K, T_\mathrm{eff}~|~ \tau, m_0)\ \ \xi (m_0)\ \ b\bigl ( \tau~|~\vec{x}, \vec{\alpha} \bigr ) . 
\end{align}
In Eq.\ref{eq:basic-prediction} the term $S_c(M_K, T_{\mathrm{eff}})$ reflects the observational selection function of the sample as specified in Eq. \ref{eq:selfunc}; $p(M_K, T_\mathrm{eff} | \tau, m_0)$ reflects the probability density of a star to have absolute magnitude $M_K$ and effective temperature $T_\mathrm{eff}$ at a given birth mass $m_0$ and age $\tau$; and $\xi (m_0)$ is the initial mass function in units of $\mathrm{mass}~[M_{\odot}]^{-1}$.  
The integral over age $\tau$ should in principle extend from 0 to $t_{Hubble}$. In practice, we consider minimum and maximum ages from $\tau_{min}$ to $\tau_{max}$, as the very youngest stars are too enshrouded to be in the sample, and stars $\tau \gtrsim 1$~Gyr will be too faint in their main-sequence phase.
In the next section we describe the single terms of Eq. \ref{eq:basic-prediction}.

\subsection{Functional forms of the model components}
\vspace{2mm}\textit{Birthrate:\ }
We specify the birthrate $b\bigl ( \tau~|~\vec{x}, \vec{\alpha} \bigr)$ as a function of time $\tau$ at a given position in the Galaxy $\vec{x}$, given parameters $\vec{\alpha}$. We want to avoid imposing a particularly restrictive functional form, as we have no reason to believe that the age distribution is particularly smooth or spatially homogeneous. Therefore, we parameterize it by step functions in time that can be different in each spatial pixel $\vec{x}$:
\begin{equation}
b\bigl ( \tau~|~\vec{x},\vec{\alpha} \bigr) = \sum_{\mathrm{age~bin}~i} \alpha_i\bigl ( \vec{x} \bigr )~ \chi_{A_i}(\tau),
\label{eq:stepfunction}
\end{equation}
where $A_i$ are age intervals, the rate coefficients $\alpha_i$ are the elements of $\vec{\alpha}$ and must satisfy $\alpha_i > 0$, and the indicator function $\chi_{A_i} = 1$ if $\tau \in A_i$ and zero otherwise.
After some experimentation to explore the trade-off between functional flexibility and the noisiness of individual bins, we adopted five bins:
\begin{align*}
A_1 &= 5-10 \, \mathrm{Myr} \\
A_2 & = 10 - 30 \, \mathrm{Myr} \\
A_3 &= 30 - 100 \, \mathrm{Myr} \\
A_4 &= 100 - 300 \, \mathrm{Myr} \\
A_5 &= 300 \, \mathrm{Myr} - 1 \, \mathrm{Gyr},
\end{align*}
spaced approximately logarithmic in age, which corresponds to a linear dependence in turn-off magnitude. 

\vspace{2mm}\noindent\textit{Initial mass function:\ }
We use a broken power law Kroupa IMF \cite{Kroupa2001, Kroupa2002} whose piecewise components $\xi(m) \approx m^{-\Gamma}$ are:
\begin{align}
    \Gamma &=  0.3 \hspace{0.5cm}  \mathrm{for}\,\,\  0.01 < \mathrm{M / M_{\odot}} < 0.08\nonumber\\
     \Gamma &= 1.8 \hspace{0.5cm}  \mathrm{for}\,\,\  0.08 < \mathrm{M / M_{\odot}} < 0.5\nonumber\\
      \Gamma &= 2.7 \hspace{0.5cm} \mathrm{for}\,\,\  0.5 < \mathrm{M / M_{\odot}} < 1\nonumber\\
       \Gamma & =2.3 \hspace{0.5cm} \mathrm{for}\,\,\  1 < \mathrm{M / M_{\odot}} < 120. \nonumber
\end{align}
This form of the IMF accounts for unresolved binaries, which are important across most of the IMF's mass spectrum \citep[e.g.][]{Moe2017}. 
Previous studies \citep{Robitaille2010, Chomiuk2011} have not used IMF forms that make this correction. The choice of the IMF matters in this context, as all such studies match the observed frequency of young stars in a finite mass range and then extrapolate to the SFR across the full range of $\xi(m)$.
 We discuss in Section \ref{sec:discussion} how our results would have been affected by using an alternate, simpler IMF model \citep[e.g.,][]{Chomiuk2011,Robitaille2010}.

\vspace{2mm}\noindent\textit{Selection function:\ }
The term $S_c$ in Eq.~\ref{eq:basic-prediction} is the selection function of the sample, i.e. a function that returns the (unitless) probability that a star is in the photometric catalogue we model, given its observed properties. In this work, we approximate the selection function described in Section \ref{sec:data} and Z21 as a function of $M_K$ and $T_\mathrm{eff}$. We presume that  $S_c(M_K, T_{\mathrm{eff}})$ is either 1 or 0. Since $M_K$ is an observable, the function $S_c$ in $M_K$ is directly linked to the selection of the observed sample. The dependence of $S_c$ on $T_{\mathrm{eff}}$ is somewhat more indirect, as it is \emph{not} a direct photometric observable, especially in the presence of reddening. In practise we only need to know $T_{\mathrm{eff}}$ well enough to eliminate RGB stars at that luminosity, which is possible even for reddened stars (Z21). The selection function $S_c$ assures that the integral in Eq.~\ref{eq:basic-prediction} has non-zero contributions only from parts of the isochrones where both $M_K$ and $T_{\mathrm{eff}}$ are within the ``selected'' range.  
In practise, we approximate the sample selection function as:
\begin{equation}
\label{eq:selfunc}
    S_c(M_K, T_{\mathrm{eff}}) = 
    \begin{cases}
    1 ~~~~ {M_K < \min\{0,M_K^{max}(\vec{x})\} }~\wedge~ T_{\mathrm{eff}} > 7000 \mathrm{K} \\
    0 ~~~~ \text{otherwise.}
    \end{cases}
\end{equation}
The condition $M_K < \min\{0,M_K^{max}(\vec{x})\}$ arises from the fact that for the most distant patches \vec{x} the apparent magnitudes $G$ of stars with absolute magnitudes $M_K=0$ may be too faint and lie in a magnitude range where \textit{Gaia} is not necessarily complete. We therefore impose a brighter cut to ensure that we can fully model the selection function.
We derive $M_K^{max}(\vec{x})$ by smoothing the observed $M_K$ distribution
with a Gaussian kernel, and taking the mode of the distribution. This reduces the number of stars in each spaxel to 550-800 stars, depending on distance (the more distant spaxels have fewer sources).

\vspace{2mm}

\noindent
\textit{Probability density of $M_K$:\ }
The overall data-model comparison requires us to predict how many hot stars of luminosity $M_K$ we would expect from isochrones for a stellar population of unit mass and age $\tau$: $p(M_K | \tau)$ (in units of $\mathrm{mag}^{-1}$). 
To compute $p(M_K | \tau)$ we first apply our selection function $S_c(M_K, T_{\mathrm{eff}})$, and marginalise (integrate) over all $T_\mathrm{eff}$ and $m_0$. We use PARSEC isochrones \citep{Bressan2012, Tang2014, Chen2015} from 1 Myr to 1 Gyr,  equally spaced by 0.1 dex in log(age/yr). Then, for each age, we sample the Kroupa IMF and assign to each star the appropriate absolute magnitude $M_K$ by finding the nearest isochrone point according to the mass of the star. Finally, we count the number of stars in bins of width $\Delta M_K = 0.25$~mag, between $M_{K, min}$ and $M_{K, max}$ and we normalise such number for the total number of stars and the bin width $\Delta M_K$. 
Therefore, we obtain the quantity $n(M_K, \tau)$ that corresponds to the integral over the initial mass and effective temperature $m_0, T_\mathrm{eff}$ of Eq. \ref{eq:basic-prediction}:
\begin{align}
\label{eq:nMk,tau}
    n(M_K, \tau) &= \int dm_0 \int dT_\mathrm{eff} \nonumber \\
& S_c(M_K, T_{\mathrm{eff}})~p(M_K, T_\mathrm{eff} ~|~ \tau,m_0)~\xi (m_0).
\end{align}

Since we do not know the individual ages of the stars in bin $\vec{x}$, we also need to integrate over 
isochrone age $\tau$ before comparing with the data:
\begin{equation}
    n(M_K ~|~ \vec{x}, \vec{
\alpha}) = \int_{\tau_{min}}^{\tau_{max}} n(M_K, \tau) b\bigl ( \tau~|~\vec{x}, \vec{\alpha} \bigr ) d\tau,
\end{equation}
which we do numerically. As the isochrones are logarithmically spaced, it proves useful to define the logarithmic age $\omega= \log\tau$ and transforming integration variables to get
\begin{equation}\label{eq:nmk}
   n\bigl (M_K~|~\vec{x}, \vec{
\alpha}) = \int_{i, \omega_{min}}^{i, \omega_{max}} n(M_K~|~10^{\omega_i})~b(10^{\omega_i})~\ln(10)e^{\ln(10)\omega_i}~d\omega
\end{equation}
where we dropped the arguments $\vec{x}$ and $\vec{\alpha}$ from $b$ for conciseness.
Now, Eq. \ref{eq:nmk} represents the predicted $M_K$-distribution at a given position $\vec{x}$, given the model assumptions about $b\bigl ( \tau~|~\vec{x},  \vec{\alpha}\bigr )$.

\subsection{The data likelihood function}
We compare the predictions from Eq. \ref{eq:nmk} to the observed magnitude distribution $\{M_{K,i}\}$ detected in the patch $\vec{x}$ by quantifying and optimising the likelihood of the data.
We assume that the stars in our data set are sampled by an inhomogeneous Poisson point process with a rate function $\lambda = n(M_{K}~|\vec{x}, \vec{\alpha})$. This leads the joint probability of the data, given the model parameters $\vec{\alpha}$ of $p(N=N_\star, \{M_{K,i}\}| \vec{\alpha} )$ (the likelihood) to be the product of the two terms specified below.
The first term sets the normalisation, and informs on the total size of the sample,
\begin{equation}\label{Eq:first_likelihood}
    \frac{\left[\int n(M_{K}~|~ \vec{x}, \vec{\alpha}) d M_{K}\right]^N} {N!} \exp{\left[-\int n(M_{K}~|~\vec{x}, \vec{\alpha}) dM_{K}\right]} 
\end{equation}
states that the probability of observing N stars in a spaxel follows a Poissonian distribution.
The second term
\begin{equation}\label{Eq:second_likelihood}
    \prod_i \frac{n(M_{K,i} ~|~ \vec{x}, \vec{\alpha})}{\int n(M_{K}~|~ \vec{x}, \vec{\alpha}) dM_K}, 
\end{equation}
specifies the form of the distribution of the data and states that the observed rate is normalised over the volume in $M_K$ space that could
have been observed within the survey selection constraints as expressed by the survey selection function \citep[e.g.][]{Bovy2012,RixBovy2013,Rix2021}.
By multiplying Eqs. \ref{Eq:first_likelihood} and \ref{Eq:second_likelihood} we obtain:
\begin{equation}\label{eq:likelihood}
    \mathcal{L} \left(~\{M_{K,i}\}~|~\vec{x},\vec{\alpha} \right) =\Biggl ( \prod_{i = 1}^{N_*} n (M_{K,i}~|~ \vec{x}, \vec{\alpha})\Biggr )~e ^{-N_{*,tot}},
\end{equation}
where the normalization factor is:
\begin{equation}
N_*^{tot}(\vec{x}) = \int_{M_{K, min}} ^{M_{K, max} = 0} n(M_K\,|\,\vec{x}, \vec{
\alpha})~dM_K.
\end{equation}
Using uniform priors between 0 and 0.1, we sample the $\alpha_i (\vec{x})$ parameters characterising the birthrate $b\bigl ( \tau~|~\vec{x}, \vec{\alpha} \bigr )$ in each spaxel 
from the posterior with \texttt{emcee} \citep{emcee}, and using respectively the 50\textit{th}, 16\textit{th}, and 84\textit{th} percentiles as the parameter best estimate and the 1$\sigma$ levels. We call  $b\bigl ( \tau~|~\vec{x}, \vec{\alpha}_{best} \bigr )$ the birthrate obtained by using the best estimates of the $\alpha_i$ parameters.

\subsection{Tests on mock data}
\begin{figure*}[h]
    \centering
    \includegraphics[width = \hsize]{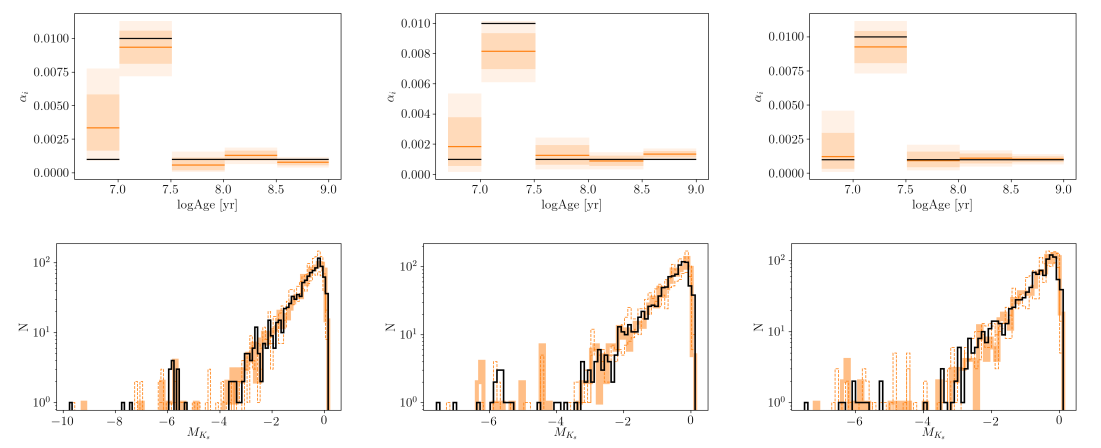}
    \caption{Model fits to mock datasets used as tests. Top row: the black lines show the true value of the birthrate $\alpha_i$ parameters. The orange lines show the estimated $\alpha_i$ parameters for three different realisations of the mock data sets generated from the same initial parameters. The orange shaded areas correspond to the $1=$ and $2-\sigma$ levels. Bottom row: the black histograms represent the magnitude distribution of three different mock data sets. The orange thick histograms represent the magnitude distribution obtained by using the best fit $\alpha_i$. The dashed histograms represent the magnitude distribution obtained by using the $16th$ and $84th$ percentiles for $\alpha_i$.}
    \label{fig:sfr_mock}
\end{figure*}
To test our model, we generate a mock magnitude distribution $n(M_K|\vec{x}, \vec{\alpha})$ (see Eq. \ref{eq:nmk}) by using a known birth rate $b\bigl ( \tau~|~\vec{x}, \vec{\alpha}_{true} \bigr ) = \sum_i \alpha_{i, true} \chi_{A_i}(\tau)$ (where we chose here $\vec{\alpha}_{true} = \vec{\alpha}_{best}$), and the same selection function, set of isochrones, and IMF as used in our model. We then 
use our fitting procedure described in the previous section to derive the parameters $\alpha_{i, \mathrm{pred}}$ and compare them with the true values $\alpha_{i, \mathrm{true}}$. To use this test in a physically meaningful range of the parameter space and to anticipate the results (Section \ref{sec:results},) the $\alpha_{i, \mathrm{true}}$ are chosen to reproduce the observed number of stars in a spaxel at a given position $\vec{x}$. 

Figure \ref{fig:sfr_mock} (top row) shows the true (black lines) and predicted values (orange lines) for $\alpha_i$, for three realisation of the same mock magnitude distribution. The shaded areas correspond to $1\sigma$ and $2\sigma$ levels. Figure \ref{fig:sfr_mock} (bottom row) shows the corresponding true mock magnitude distribution (black histogram) and the predicted magnitude distribution for the best fit $\alpha_i$ value (thick solid orange histogram) and the $1\sigma$ values (dashed thin histograms).
The predicted parameters are compatible within 1-2$\sigma$ with the true values and they are always consistent with each other. The increase in birth rate in the second age interval (10-30 Myr) is always correctly retrieved. The difference between the estimated values of the parameters most likely depends on the different sampling of the bright end of the magnitude distribution ($M_K < -4$~mag), which is also where the true magnitude distribution differs the most from that estimated using the predicted parameters. This is due to the very low number of stars predicted (and observed) at extremely high luminosity.


\section{Results}\label{sec:results}
\begin{figure}[h]
    \centering
    \includegraphics[width = \hsize]{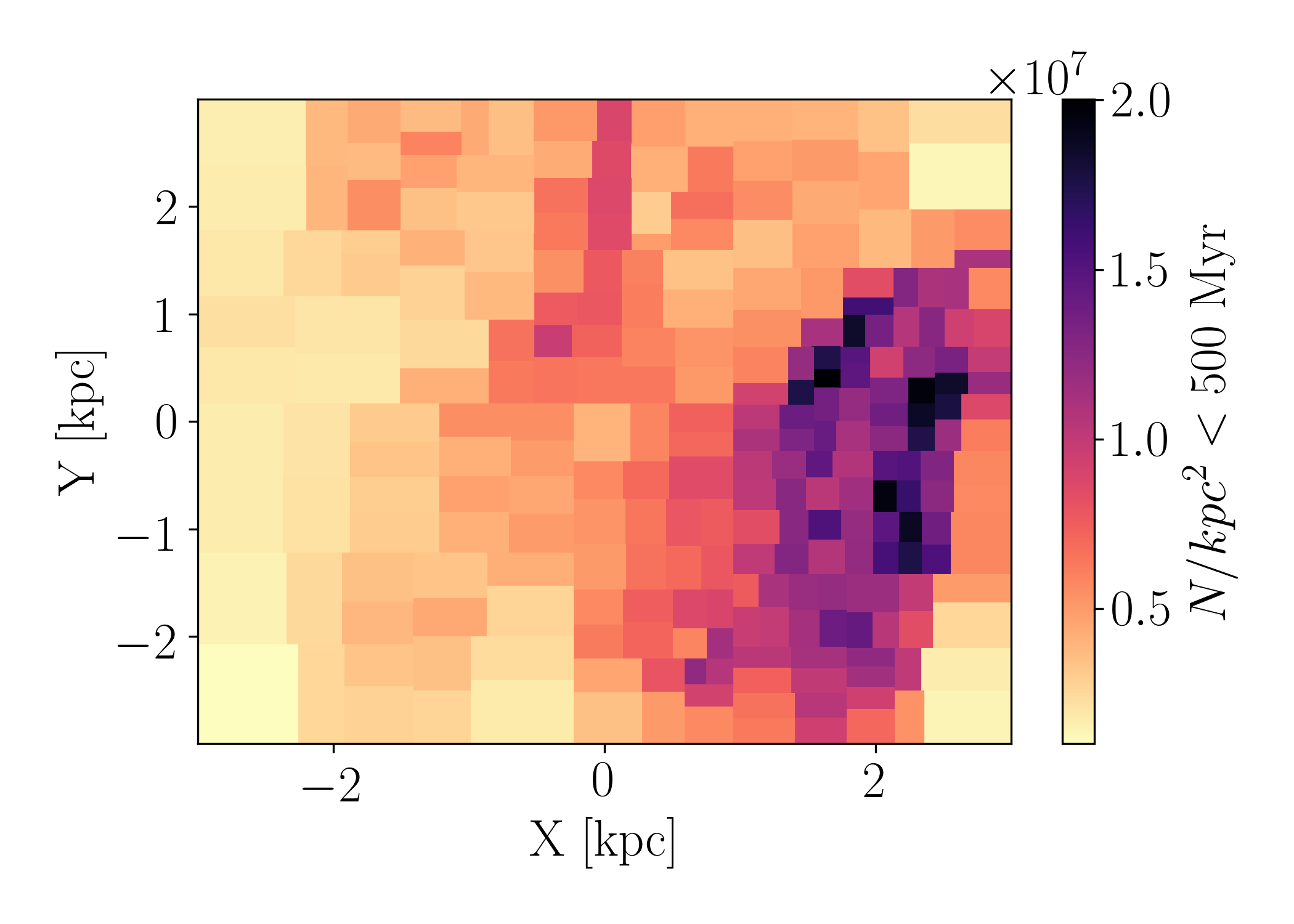}
    \caption{Density distribution of the stars younger than $500$ Myr in the Galacic plane, as derived from Eq. \ref{eq:N}, i.e. by using the best estimates for the birthrate coefficients $\alpha_{i}$ in each spaxel $j$.}
    \label{fig:stars_kpc_squared_all}
\end{figure}

In this Section we present the results obtained by fitting our model to the data presented in Section \ref{sec:data}. We compute the values and uncertainties for the coefficients $\alpha_{i,j}$ that imply $b(\tau_i~|~\vec{x}_j, \vec{\alpha}_j)$ for all the spaxels $\vec{x}_j$ in the map of Fig. \ref{fig:density} (right) by sampling the likelihood of Eq. \ref{eq:likelihood}. Example values of the $\alpha_i$ parameters and their uncertainties are given in Table \ref{table:results} and will be available on CDS.
\begin{table*}
\caption{Values and uncertainties for three spaxels, and corresponding coordinates. The full result table will be available upon request and on CDS.}
\begin{center}
\begin{tabular}{ c|c|c|c|c|c|c }
& & & & & & \\
 $x_{\mathrm{min}}$ -- $x_{\mathrm{max}} [kpc]$ & $y_{\mathrm{min}}$ --  $y_{\mathrm{max}}$ [kpc] & 
 $\alpha_{\alpha_{16}}^{\alpha_{84}}$ & $\alpha_{\alpha_{16}}^{\alpha_{84}}$ & $\alpha_{\alpha_{16}}^{\alpha_{84}}$ & $\alpha_{\alpha_{16}}^{\alpha_{84}}$ & $\alpha_{\alpha_{16}}^{\alpha_{84}}$ \\
 & &  5-10 Myr & 10-30 Myr & 30-100 Myr & 100-300 Myr & 300 Myr - 1 Gyr \\
 \hline
 & & & & & &  \\
 2.23--2.99 & 2.57--2.99 & $0.009_{0.008}^{0.01}$ & $0.0006_{0.0002}^{0.0016}$ & $0.00025_{0.00007}^{0.0006}$& $0.0026_{0.0024}^{0.0029}$ & $0.00009_{0.00002}^{0.0002}$\\
 & & & & & & \\
 2.23--2.99 &  1.96--2.57 & $0.011_{0.009}^{0.013}$ &  $0.0022_{0.0006}^{0.0046}$ & $0.00033_{0.00009}^{0.0008}$ & $0.0010_{0.0006}^{0.0014}$  &  $0.0015_{0.001} ^{0.002}$ \\
 & & & & & & \\
 1.86--2.23 & 2.49--2.99 & $0.008_{0.007}^{0.009}$ & $0.0018_{0.0007}^{0.0034}$ & $0.0007_{0.0002}^{0.0014}$ & $0.0019_{0.0016}^{0.0022}$ & $0.00026_{0.00009} ^{0.00052}$
\end{tabular}
\end{center}
\label{table:results}
\end{table*}
These results are essentially number- or mass-density maps of mono-age stellar populations (in five age bins) across our surrounding 6kpc$\,\times\,$6kpc area of the Galactic disc (Figs. \ref{fig:stars_kpc_squared_all} and \ref{fig:stars_kpc_squared}), as well as the overall radial gradients they show (Figs. \ref{fig:radial_sfr} and \ref{fig:radial_sfr_time}). When summing over the entire area we obtain an estimate of the integrated age distribution of the stars (Fig.\ref{fig:total_sfh}), which reflects the formation rate of the stars that are \emph{now} in this portion of the disc. 

\subsection{Mono-age stellar density maps of the Galactic disc}
Figure~\ref{fig:stars_kpc_squared_all}~shows the number surface density of stars $n(\Delta\tau~|~\vec{x}, \vec{\alpha})$, younger than 500~Myr, defined as:
\begin{equation}\label{eq:N}
    n(\Delta\tau~|~\vec{x}, \vec{\alpha}) \equiv  \frac{1}{\mathrm{Area}(\vec{x})}\int_{\tau_{min,i}}^{\tau_{max,i}} b_ {best}(\tau~|~\vec{x}, \vec{\alpha}_{best})~ d\tau , 
\end{equation}
where the integral of $b(\tau~|~\vec{x}, \vec{\alpha}_{best})$ over time gives the number of stars formed
between $\tau_{min} < \tau < \tau_{max}$ and $\mathrm{Area}(\vec{x})$ is the spaxel area in kpc$^2$.
The density distribution in the Galactic plane of sources younger than 500 Myr is qualitatively very similar to that shown in Fig.  \ref{fig:density}, lending credence to our results. The \emph{total} number of stars in each spaxel is different from the number of \emph{observed} stars per spaxel because with Eq. \ref{eq:N} we predict the total number of stars at \emph{all} masses, not only those included in the filtered sample from Z21. These maps can be converted into maps of the (hypothetical, initial) stellar surface mass density of stars in that age bin, simply by multiplying with the initial mean mass per star, $\langle M_*\rangle = 0.22\, \mathrm{M_{\odot}}$ for our Kroupa IMF.

\begin{figure*}
    \centering
    \includegraphics[width = \hsize]{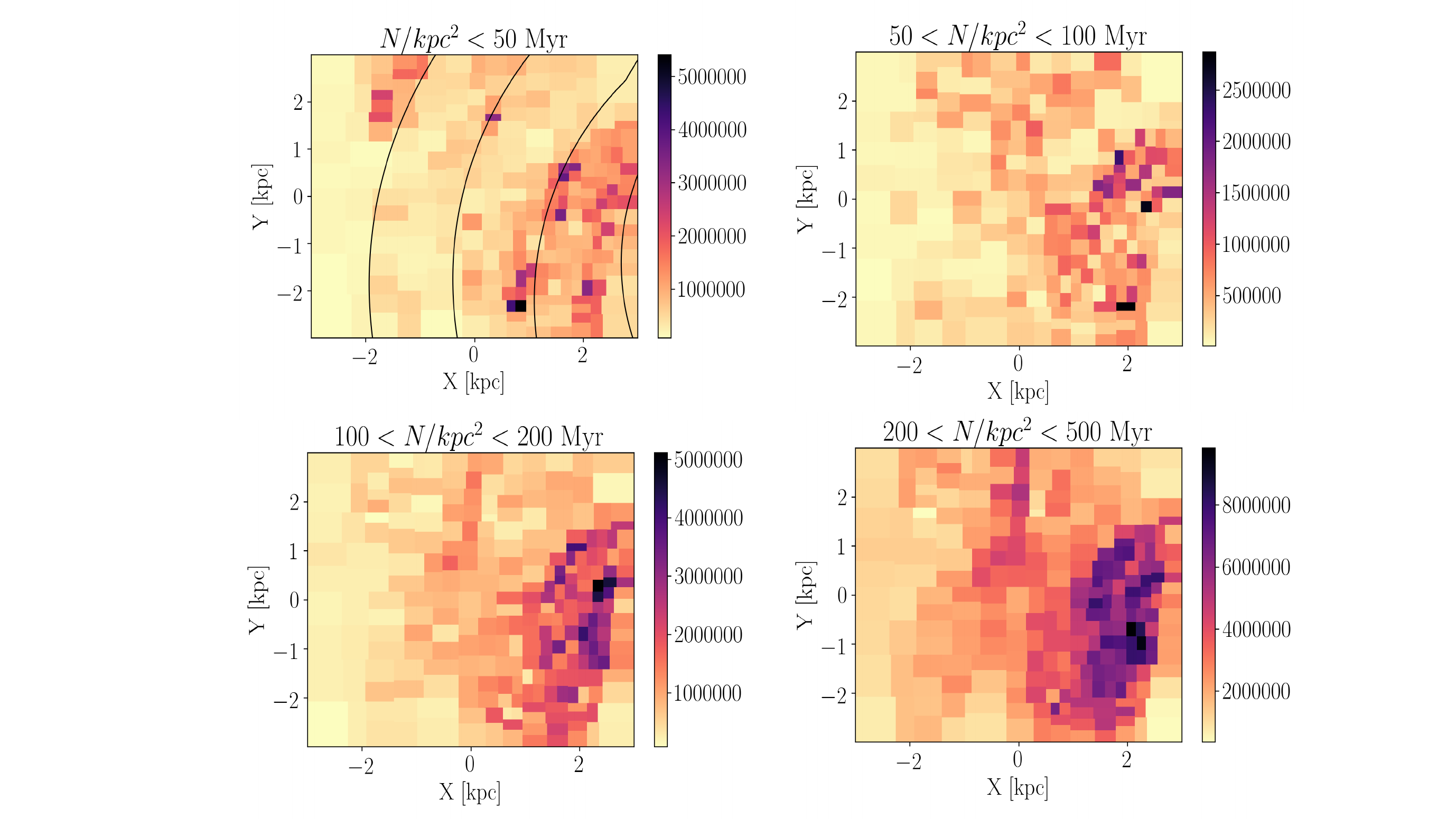}
    \caption{Density distribution of stars with ages $5 < t < 50$ Myr (top left), $50 < t < 100$ Myr (top right), $100 < t < 200$ Myr (bottom left), and $200 < t < 500$ Myr (bottom right). The density distribution changes as a function of time, becoming gradually smoother. Over-arching structures corresponding to spiral arms are visible for all age intervals. The solid black lines indicate the location of the spiral arms from 
    \cite{Reid2019}.}
    \label{fig:stars_kpc_squared}
\end{figure*}
The density distribution of the sample can be further investigated by splitting the stars into age bins, with ages $5 < \tau< 50$ Myr (Fig. \ref{fig:stars_kpc_squared}, top left), $50 < \tau < 100$ Myr (Fig. \ref{fig:stars_kpc_squared}, top right),  $100  < \tau < 200$ Myr (Fig. \ref{fig:stars_kpc_squared}, bottom left), and $200 < \tau < 500$ Myr (Fig. \ref{fig:stars_kpc_squared}, bottom right). The appearance of these maps changes as a function of age. In the youngest age interval ($5 < \tau< 50$ Myr) the density distribution traces massive star-forming regions, which are visible as prominent overdense clumps in Fig. \ref{fig:density} and Fig. \ref{fig:stars_kpc_squared_all} \citep[see also][]{Zari2021, Poggio2021}. The distributions of stars in the less young age intervals appear smoother and not show dense clumps, certainly not beyond the age of one dynamical dime scale or Galactocentric rotation, $\sim 200$\, Myr. However, arch-like or spiral-like structures are visible for all age groups. The strength of spiral arms is quantified in Fig. \ref{fig:sfr_azimuth}.

\subsection{Radial trends of the young star density}
For potential comparison with in particular external galaxies \citep{Gonzalez2016} we now look at the dependence of the young star density on Galactocentric radius $R_{GC}$, averaging over the azimuthal angle that lies inside our 6kpc$\,\times\,$6kpc box, simply summing over all spaxels in a certain $R_{GC}$ range:
\begin{equation}
    n_R(\tau~|~R_{GC})\equiv \  \frac{1}{S} \sum_{\mathrm{spaxels~at~R_{GC}}} b( \tau~|~\vec{x}, \vec{\alpha}_{best}),
\end{equation}
where $S$ is the total area of these spaxels. 

\begin{figure}
    \centering
    \includegraphics[width = \hsize]{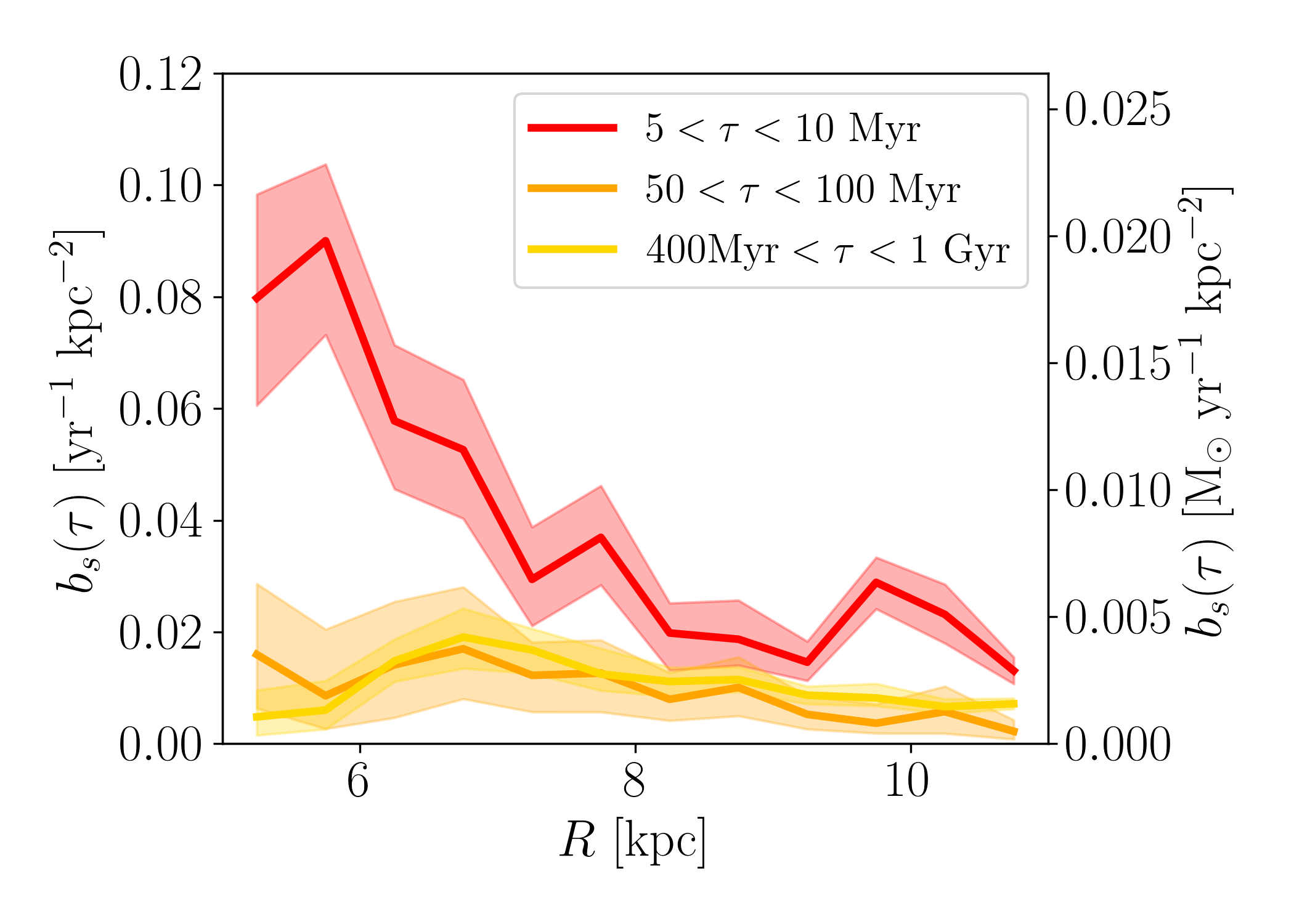}  
    \caption{Distribution of star formation rate surface density as a function of Galactocentric radius. The thick lines correspond to the average star formation rate surface density for different age intervals (red: $5 < \tau < 10$~Myr; orange: $50<\tau<100$~Myr; yellow: $400 <$~Myr~$\tau < 1$~Gyr). The shaded areas correspond to  the $16th$ and $84th$ percentiles. The left Y-axis shows the star formation rate expressed as the total number of stars born; the right Y-axis the more conventional amount of stellar mass formed, which requires the adoption of an IMF \citep[here][]{Kroupa2001}.}
    \label{fig:radial_sfr}
\end{figure}

Figure \ref{fig:radial_sfr} shows the radial distribution of $n_R(\tau_i~|~R_{GC})$, for three different age intervals $\tau_i$: $5 < \tau < 10 $ Myr, $50 < \tau < 100$ Myr, and $400 \, \mathrm{Myr} < \tau < 1 $~Gyr.  While the older age bins only show a shallow gradient, the radial density gradient among stars $\tau<10$\,Myr is quite steep.

\begin{figure}
    \centering
    \includegraphics[width = \hsize]{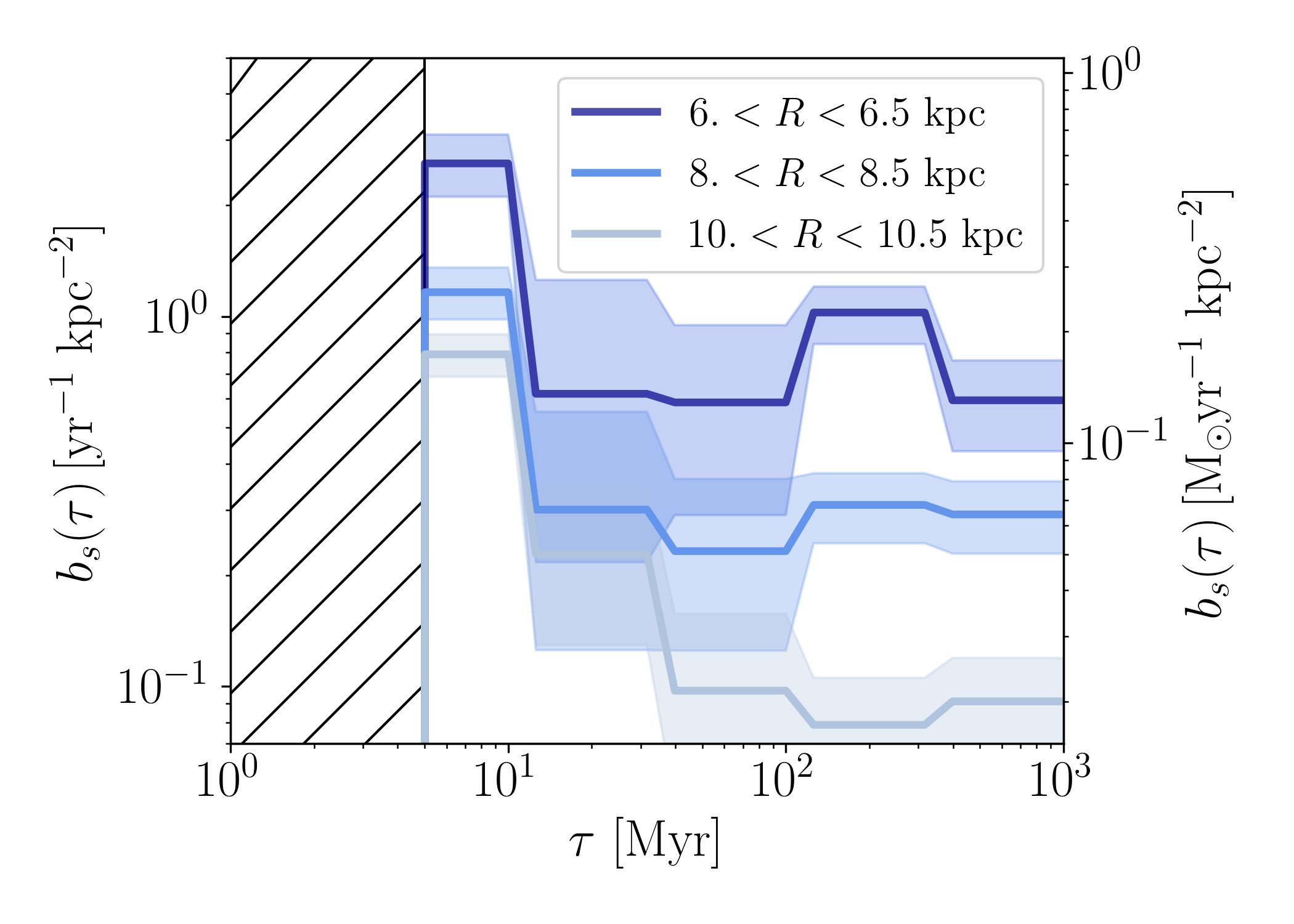} 
    \caption{Star formation rate density in different Galactocentric radii, as a function of time. The left and right Y-axes show the SFR expressed as the total number of stars born and as the stellar mass formed, respectively. The thick lines correspond to the average star formation rate surface density for three intervals in Galactocentric radius. The shades areas correspond to the $16th$ and $84th$ percentiles.  Ages younger than 5 Myr are not included in our analysis, and thus the corresponding area in the plot is hatched. }
    \label{fig:radial_sfr_time}
\end{figure}

Figure \ref{fig:radial_sfr_time} shows the azimuthally averaged mean birth rate parameter $n_R(\tau_i|R)$ for different Galactocentric radii. The different Galactocentric annuli do not only vary significantly in their normalization, but also their relative age distributions. While \emph{all} show a distinct increase at the youngest ages, this increase is fractionally largest well outside the solar radius. The age distribution well inside the solar radius may have had a second peak about 300~Myr ago. 

\subsection{Effective overall star formation rate}
As the final step in the presentation of our results we show the age distribution when summed over entire $6\times6~$~kpc$^2$ area.
We define the total "effective" birth rate $b_{tot}$ as 
\begin{equation}
    b_{tot}(\tau) = \sum_{all~spaxels~\vec{x}} b(\tau\,|\,\vec{x}, \vec{\alpha}_{best})
\end{equation}
by summing over all the $j$ spaxels in the map (and \emph{not} normalising by the area).
We shall discuss below to what extent this age distribution reflects the overall \emph{star formation rate} across the Galactic disc. 

\begin{figure}
    \centering
    \includegraphics[width = \hsize]{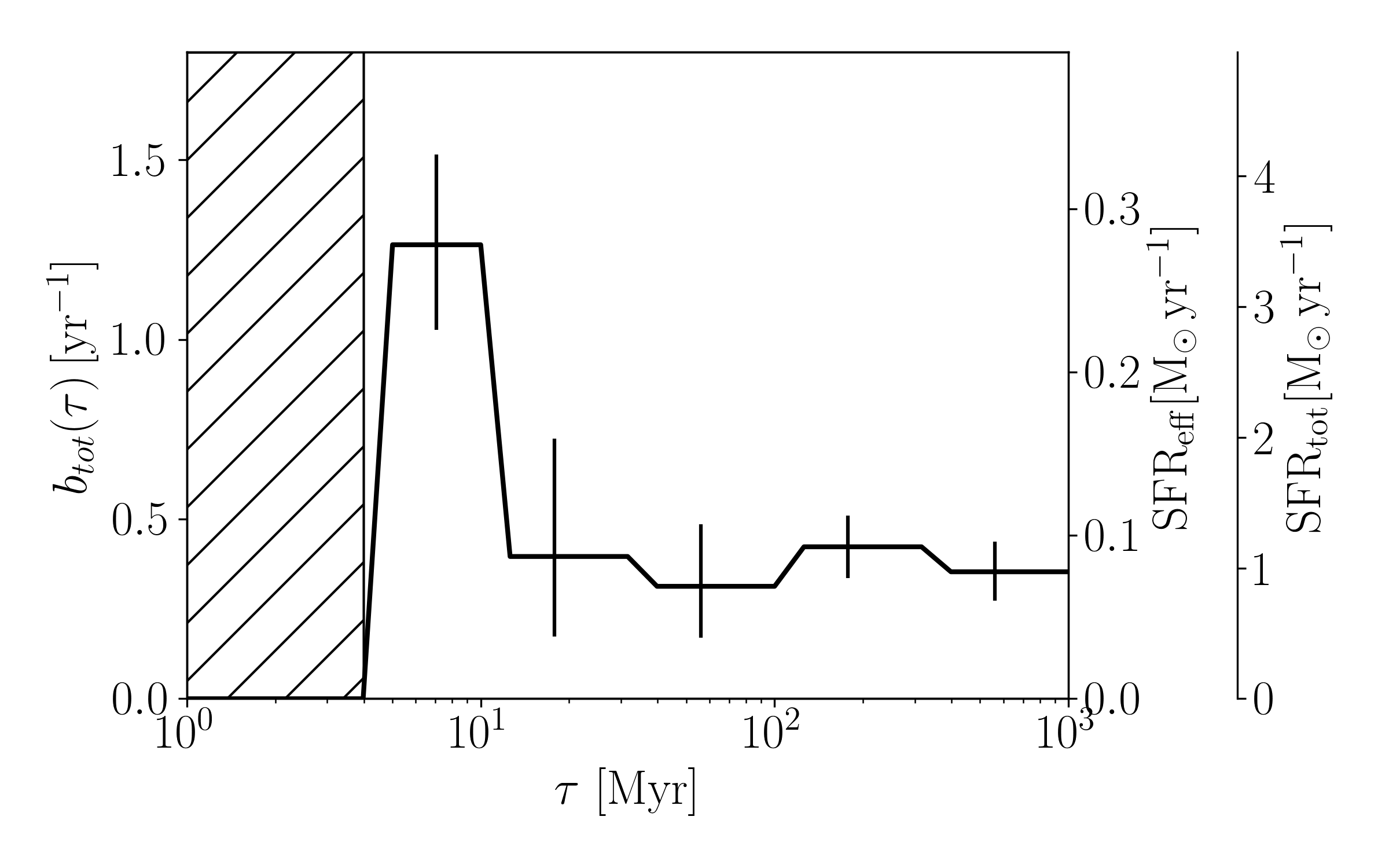}
    \caption{Effective star formation history of the sample, in units of $\mathrm{yr^{-1}}$ (left y-axis) and $\mathrm{M_{\odot}\, yr^{-1}}$ (first right y-axis). The second right y-axis corresponds to the total SFR obtained by multiplying by the ratio $f =\mathrm{SFR_{5-11}/SFR_{tot}}$ (see Section \ref{sec:discussion}).  Vertical bars correspond to errors on the parameters. Ages younger than 5 Myr are not included in our analysis, and thus the corresponding area in the plot is hatched.}
    \label{fig:total_sfh}
\end{figure}

Units of $b_{tot}(\tau)$ are \#~of stars/year, which we convert to the more conventional $\mathrm{M_{\odot} \, yr^{-1}}$ by multiplying by the mean stellar mass derived from our assumed IMF $<M> = 0.22 \, \mathrm{M_{\odot}}$. 
This is shown in Fig. \ref{fig:total_sfh}, where we show $b_{tot}(\tau)$ on the left Y-axis, and $\mathrm{SFR_{eff}}$ on the first right Y-axis. Both values refer to the entire $6\times6~$~kpc$^2$ area,  a much larger area than previous quantitative studies for this age range, but of course not the ``entire disc''.

Fig. \ref{fig:total_sfh} shows that the age distribution of stars is approximately uniform between 20~Myr and 1 Gyr at 0.4/yr (or 0.08 M$_\odot$/year. Remarkably, the effective star formation rate seems to have been three times higher in the last $\sim 10-20$~Myr.
It appears that the Galactic disc, or at least the large patch we consider here, has just lit up in young stars.

\section{Discussion and Summary}\label{sec:discussion}
In the previous sections we have estimated the recent star formation history of the extended solar neighbourhood and found that it has recently increased by a factor of $\sim$3. We now estimate the global star formation rate of the Galactic disc and we put this result into context by comparing  with previous literature studies. We then use simulations to quantify to what extent global conclusions can be extrapolated from local measurements.

\subsection{Comparison to Existing Work}
Assuming that the fraction of the disc that we are probing is representative of the entire Milky Way (see Section \ref{sec:tng}), we can derive the ``total'' SFR of the Milky Way. To do so, we consider the  youngest bin, $\tau\lesssim $~10~Myr,  as the best approximation to the ``current SFR'' and we extend $\mathrm{SFR_{eff, 5-10 Myr}}$ to the disc. 
Since the surface density of star formation decreases strongly as a function of radius, we presume that the star formation rate surface density is well described by an exponential radial profile (Eq. \ref{eq:sfrtot}) within 4.5-15 kpc, with a scale length of $L=3.5$ kpc \citep[][see Appendix B.2]{Zari2021}:
\begin{equation}\label{eq:sfrtot}
    \mathrm{SFR}_{tot} = 2\pi \int_{R_{min}}^{R_{max}} S_0 e^{-r/L} dr.
\end{equation}
We chose $R_{min} = 4.5 \, \mathrm{kpc}$ and $R_{max} = 15 \, \mathrm{kpc}$  following \cite{Reed2005} and \cite{Davies2011} and thus exclude the central region of the Galaxy. 
In the radial range covered by our data,  $\sim5$ and $\sim11$kpc from the Galactic centre, the total SFR is:
\begin{equation}
    \mathrm{SFR}_{5-11~\mathrm{kpc}} =  \Delta\phi \int_{5}^{11} S_0 e^{-r/L} dr, 
\end{equation}
where $\Delta \phi \approx 0.7$~rad corresponds to the range of azimuth angles covered. The ratio $f = \mathrm{SFR_{5-11}} / \mathrm{SFR_{tot}} \approx 8\%$ reflects the fraction of the MW disc 
considered in this study. We thus find that the total SFR of the Milky Way implied by this model is $3.3_{-0.6}^{+0.7}\, \mathrm{M_{\odot}/yr}$ (see Figure \ref{fig:total_sfh}).

Recent analogous estimates of the star formation rate of the Milky Way are between $1.65 \pm 0.19 \, \mathrm{M_{\odot} \, yr^{-1}}$ \citep{Licquia2015} and  $1.9 \pm 0.4 \, \mathrm{M_{\odot} \, yr^{-1}}$ \citep{Chomiuk2011}, silimar to but lower than the value we find. Reasons for this discrepancy may reflect differences in the assumed IMF, the spatial coverage, and the age of the tracers used; these are discussed below. 

\vspace{2mm}
\noindent
\textit{Assumed IMF.} In their Section 3, \cite{Chomiuk2011} review the different methods used to estimate the Milky Way SFR. Such methods rely on Lyman continuum photon rates \citep[e.g.][]{Mezger1978,Smith1978,McKee1997, Murray2010}, infra-red emission and YSO counts \citep{Robitaille2010, Davies2011}, or massive stars counts and supernova rates \citep{Reed2005, Diehl2006}. In their study, \cite{Chomiuk2011} \citep[and thus also][]{Licquia2015} normalised previous results to the same IMF, adopting as reference the IMF from IMF \cite[][KW03]{Kroupa2003}: $\xi(m) \propto m^{-\Gamma}$ with $\Gamma = 1.3$ for masses between 0.1 and 0.5 $\mathrm{M_{\odot}}$ and $\Gamma = 2.3$ for masses between 0.5 and 100 $\mathrm{M_{\odot}}$. 
Since our sample is predominately composed of massive stars of which most will be in unresolved binary (or multiple) systems \citep[see e.g.][]{Sana2014}, it seems appropriate to adopt an IMF that is explicitly corrected for unresolved binaries \citep{Kroupa2001, Kroupa2002} reported in Section \ref{sec:model}.  The choice of the IMF impacts, of course, the relation between the observed and modelled luminosity function $n(M_K ~|~ \tau)$ and the implied ``underlying'' total stellar mass (Eqs. \ref{eq:nMk,tau} \& \ref{eq:nmk}). 
In our formulation of the problem, this affects the inference of the birthrate parameter $\alpha_i$, and the mean stellar mass that is used to convert the SFR to units of $\mathrm{M_{\odot}}\, yr^{-1}$, which for the KW03 IMF is $<M> = 0.63 \, \mathrm{M_{\odot}}$. To quantify the impact of the choice of IMF on our results, we re-run the analysis using KW03 IMF. This yields an analogous estimate for the total SFR of $~2.3 \pm 0.4\, \mathrm{M_{\odot}}\, yr^{-1}$, which is consistent with the estimates of \cite{Chomiuk2011} and slightly higher than \cite{Licquia2015}. 

\vspace{2mm}
\noindent
\textit{Space distribution.} There are strong assumptions behind the space distribution 
of different tracers used to study the global SFR or the Milky Way (including ours, see Section \ref{sec:tng}).  For instance, the distance to HII regions are biased against distant sources, and studies that employ HII emission to compute the Galactic SFR usually account for this by doubling the luminosity of sources in our half of the Galaxy \citep{Mezger1978, Smith1978, McKee1997, Murray2010}. In other cases distances are not available, for example the space distribution of YSOs assumed by \cite{Robitaille2010} (see their Fig. 2) is quite different than what we used in this work, which is instead more similar to what proposed by \cite{Reed2005} and \cite{Davies2011}. 

\vspace{2mm}
\noindent
\textit{Mean Age of SFR Tracers.}
Finally, the lifetimes of the tracers used in other studies are different.  \cite{Robitaille2010}'s estimate ($0.68-1.45 \, \mathrm{M_{\odot} yr^{-1}}$) assume a YSO lifetime of $\approx 2$~Myr, which is younger than the minimum age considered here (5 Myr). \cite{Murray2010} use a lifetime for HII regions of 3.7 Myr following \cite{McKee1997}, slightly longer than the 3 Myr used by \cite{Mezger1978}, and again younger than our minimum age. \cite{Reed2005} assigns lifetimes from $\approx 3.5$~Myr to $\approx 20$~Myr based on the luminosities of his sample of O- and B-type stars. 
We note that our estimate of the global effective star formation rate is already $3 \times$ lower for stars that are 20 - 50~Myr old. 

The recent increase in effective SFR that we find is qualitatively similar to that found by \cite{RuizLara2020}. \cite{RuizLara2020} modelled \textit{Gaia} DR2 observed colour-magnitude diagrams to infer the star formation history within 2 kpc around the Sun and found three star formation enhancements that occurred $\sim$5.7, 1.9, and 1.0 Gyr ago and a hint of a fourth possible star formation burst spanning the last 70 Myr. \cite{RuizLara2020} did not express their results in $M_{\odot}$/yr, complicating a direct comparison.
They proposed that such enhancements are due to recurrent interactions between the Milky Way and the Sagittarius dwarf galaxy \citep{Ibata1994, Laporte2018}. We confirm with higher confidence the recent enhancement in star formation activity of the Galactic disc and we constrain it to the last 10 Myr. 
Given that we find distinct temporal variations in the Galaxy's effective star formation rate over the last 100\,Myr (Fig. \ref{fig:total_sfh}), matching the exact age distribution of the tracers would be crucial for a more quantitative comparison with \citet{Cignoni2006} and \citet{RuizLara2020}.

Figure \ref{fig:stars_kpc_squared} shows that the location of the density enhancements broadly coincides between the youngest and older stellar age intervals. This suggests that the spiral pattern is associated not only with young star-forming regions but also with the underlying mass density \citep{Eskridge2002, Binney2008}, at least for $\tau < 500$ Myr. Consequently, Fig. \ref{fig:stars_kpc_squared} represents evidence that those spiral arms that show up in all our monoage maps are indeed stellar mass overdensities.  Spiral arms that are seen in stellar populations of different ages are also observed in external galaxies. For instance, \cite{Meidt2021} compare the distribution of molecular gas (tracing star formation) and $3.6 \, \mu m$ emission (tracing old disc populations) across 67 star-forming galaxies with different morphological properties, often finding coincidence. 
However, not all spiral arms in the Milky Way disc, inferred e.g. from masers, are reflected in stellar mass overdensities in our maps among stars with $\tau \lesssim 0.5$~Gyrs (see top left panel of Fig. \ref{fig:stars_kpc_squared}).

\subsection{How closely does the best-fit age distribution reflect the overall disc SFR?}\label{sec:tng}
The area sampled by our OBA sample is only $\approx$8\% of the entire Galactic disc. Large-scale spatial variations in the star formation rate density or stellar ages, such as large-scale spiral arms, could bias the generalisation from this sample to the entire Galactic disc. To test how much variance could be expected, we examine the final snapshot of cosmological simulations of Milky Way-like galaxies. We use the TNG50 cosmological simulation \citep{Pillepich2019, nelson_2019b}, whose numerical resolution reaches that of zoom-in simulations (reaching a baryonic mass of $8.5\times10^4M_\odot$). IN TNG50, we select Milky Way-like galaxies based on their stellar mass, diskyness and morphology. Specifically, we require:
\begin{itemize}
    \item $10^{10.5} \leq M_\star/M_\odot \leq 10^{11.2} $,
    \item over  40\% of stars are on near circular orbits \citep[with circularities $\geq 0.7$,][]{genel_2015},
    \item have a central bar,
    \item have a non-zero star formation rate at $z=0$,
\end{itemize}
leading to $N = 78$ galaxies. We centre and align a Cartesian coordinate system with the vertical axis aligned with the angular momentum of the disc. To mimic our observed sample, we place a solar neighbourhood at $(R, z, \varphi) = (8, 0, \varphi)$ and select stars in a cylinder centred on the Sun with a 3~kpc radius. To investigate how the location $\varphi$ of the Sun influences the results, we vary the azimuth of the Sun and measure the local star formation history (from the local age distribution), thereby estimating the amplitude biases in local datasets. 

\begin{figure}
   \includegraphics[width = \hsize]{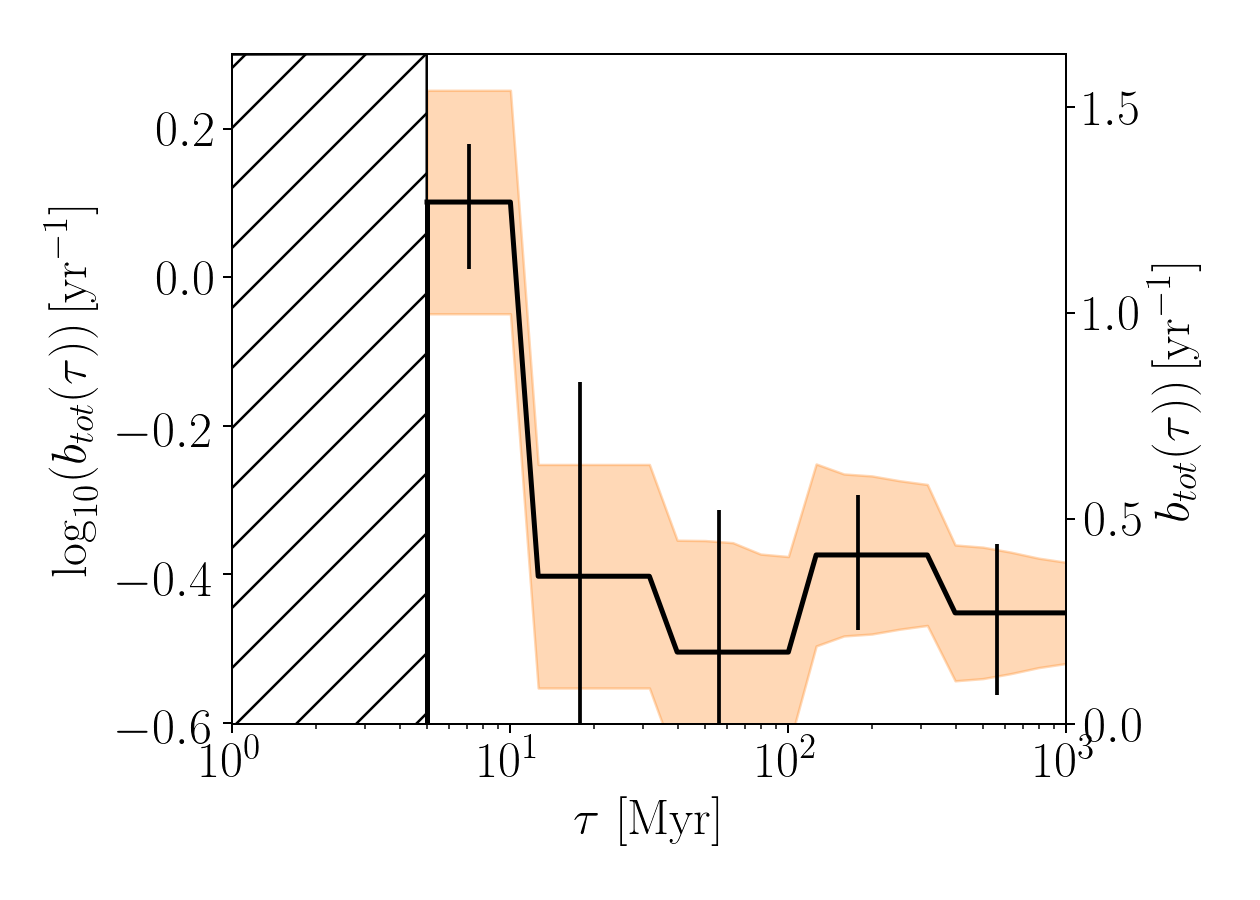}
    \caption{Logarithm of the effective star formation history of the sample shown in Fig. \ref{fig:total_sfh}. The orange bands show the expected azimuthal variations of the effective star formation history as derived by analysing Milky Way analogues in TNG50. }
   \label{fig:compare_tng50}
\end{figure}

Sampling all galaxies, and all azimuthal angles for the position of the hypothetical sun, we derive the distribution of
$$\log_{10} \left( \frac{\dot{M}_\star/dS(\mathrm{patch})}{\dot{M}_\star/dS(\mathrm{annulus})} \right), $$ where the denominator is the number of stars per age interval in the pacth, and the denominator the same quantity in the entire annulus. 
We find that azimuthal variations of the recent ``effective" SFR in a patch of R=3~kpc reflect the global average typically within factor of $\sim 1.5$. These variations are shown in Fig.\ref{fig:compare_tng50} are orange bands, reflecting the \emph{conceptual uncertainties} of extrapolating the age distribution measured within our volume to a global star formation rate of the Galactic disc.
This figure shows that these uncertainties are considerable. Nonetheless, the increase in the frequency of very young stars we find reflects a global recent increase in the Galaxy star formation rate. It is unlikely that the Sun has just moved into a patch of higher star formation activity of late.

\subsection{Summary}\label{sec:conclusions}
In this study we have modelled the observed $K-$band absolute magnitude distribution $M_K$ of the sample of massive young stars presented in \cite{Zari2021} to infer the age distribution of stars within $\sim 3$~kpc from the Sun and the effective star formation rate (SFR) of the Milky Way.

We divided our volume into spaxels of different sizes and assumed that in each spaxel the $M_K$ distribution depended only on the birth rate. We parametrised the birthrate with a step function with five age intervals to avoid imposing a particularly restrictive functional form, and we derived the parameters characterising the birthrate by comparing our predicted magnitude distribution with the data magnitude distribution by quantifying the likelihood of the data.

We derived the density distribution of stars of different ages in the Galactic plane. We found that stars younger than $<50$~Myr are clustered in compact groups that trace Galactic spiral arms. Over-densities corresponding to spiral arms are however also visible in the density distribution of older populations.

We derived the star formation history of our sample over the last Gyr by estimating the effective birthrates (i.e. the birthrate that would lead to the observed magnitude distribution) for all the spaxels in our volume, and adding them over all the spaxels. 

We found that the age distribution (or total birthrate) is almost uniform between 20 Myr and 1 Gyr, but has increased by almost a factor or 3 in the last $\sim$~10-20 Myr. Such an increase can be related to the large number of massive star-forming regions within our volume. By using TNG50 simulations of Milky Way analogue galaxies, we derived that azimuthal SFR variations are considerable but smaller than the recent SFR increase. Based on the assumption that the local value could be extended to the entire Milky Way disk, we estimated the \emph{current} (last $\sim 10$~Myr) SFR of the entire Milky Way disk to be $3.3_{-0.6}^{+0.7}\, \mathrm{M_{\odot}/yr}$.

\begin{acknowledgements}
NF was supported by the Natural Sciences and Engineering Research
Council of Canada (NSERC), [funding reference number CITA 490888-16] through the CITA postdoctoral fellowship and acknowledges partial support from a Arts \& Sciences Postdoctoral Fellowship at the University of Toronto.\\
This work has made use of data from the European Space Agency (ESA) mission {\it Gaia} (\url{https://www.cosmos.esa.int/gaia}), processed by the {\it Gaia} Data Processing and Analysis Consortium (DPAC,
\url{https://www.cosmos.esa.int/web/gaia/dpac/consortium}). Funding for the DPAC
has been provided by national institutions, in particular the institutions
participating in the {\it Gaia} Multilateral Agreement.\\
This publication makes use of data products from the Two Micron All Sky Survey, which is a joint project of the University of Massachusetts and the Infrared Processing and Analysis Center/California Institute of Technology, funded by the National Aeronautics and Space Administration and the National Science Foundation.\\
This research made use of Astropy, \citep{Astropy2013, Astropy2018}, matplotlib \citep{matplotlib}, numpy \citep{harris2020array}, scipy \citep{2020SciPy-NMeth}. This work would not have been possible without the countless hours put in by members of the open-source community all over the world.
\end{acknowledgements}

\appendix
\section{Azimuthal birthrate variations}
\begin{figure*}
    \centering
    \includegraphics[width = \hsize]{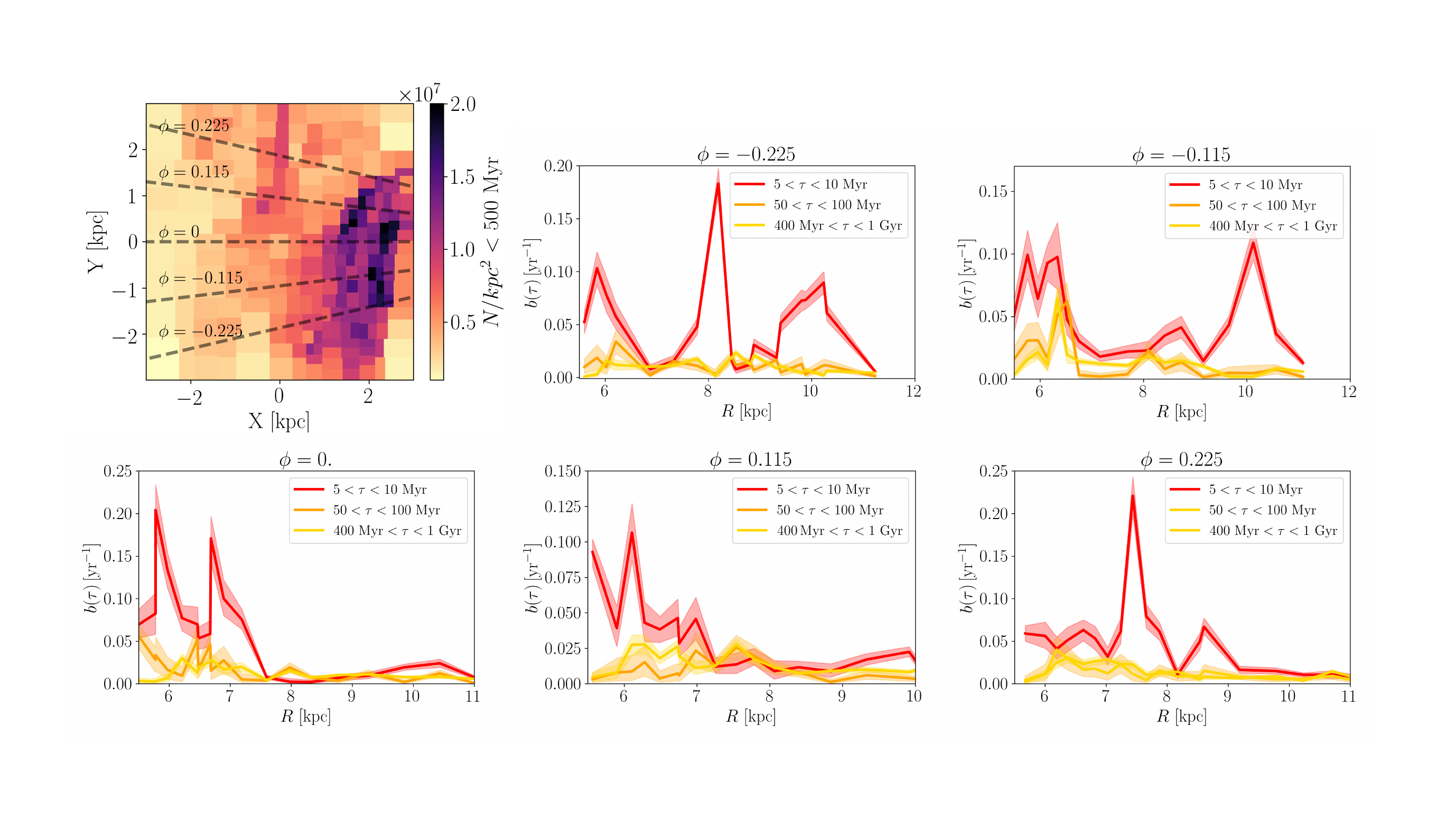}
    \caption{Top left: same as Fig. \ref{fig:stars_kpc_squared_all}, the dashed lines correspond to different azimuths. Top center, right and bottom: birthrate as a function of Galactocentric radius for three different age intervals at five different azimuths (as indicated in the panels). The colour scheme is the same as in Fig. \ref{fig:radial_sfr}. The thick lines correspond to the average birthrate, and the shaded areas correspond to the $16th$ and $84th$ percentiles.}
    \label{fig:sfr_azimuth}
\end{figure*}

To quantify the birthrate spatial variations, we consider five azimuths and we study the variations of $b_{best}(\tau | \vec{x}, \vec{\alpha})$ as a function of Galactocentric radius for the same age intervals as in Fig. \ref{fig:radial_sfr}. Figure \ref{fig:sfr_azimuth} shows that the sharp peaks in birthrate for $5< \tau<10$~Myr coincide with the locations of spiral arm segments, while for the older age ranges the birthrate is lower and does not correlate strongly with the position of spiral arms.

\bibliographystyle{aa} 
\bibliography{bibliography.bib}

\end{document}